\documentclass[sigconf,nonacm]{acmart}

\usepackage{mathptmx}

\usepackage{soul}\setuldepth{article}

\usepackage{natbib}
\usepackage{amsmath,amssymb,amsfonts}
\usepackage{algorithmic}
\usepackage{graphicx}
\usepackage{textcomp}
\usepackage{xcolor}
\def\BibTeX{{\rm B\kern-.05em{\sc i\kern-.025em b}\kern-.08em
    T\kern-.1667em\lower.7ex\hbox{E}\kern-.125emX}}

\usepackage{verbatim}

\usepackage{url}

\usepackage{breakurl}
\usepackage{tikz}  
\usetikzlibrary{decorations.markings,decorations.shapes,decorations,arrows,automata,backgrounds,petri,shapes.geometric}  
\usepackage{multirow}
\usepackage{array}
\usepackage[colorinlistoftodos]{todonotes}

\newcommand{\rh}[1]{{\color{orange} #1 }}

\setstcolor{red} 

\AtBeginDocument{%
  \providecommand\BibTeX{{%
    Bib\TeX}}}

\hypersetup{bookmarksopen=true}
\begin{document}



\title[Assessing the Solid Protocol
 in Relation to Security \& Privacy Obligations]{
Assessing the Solid Protocol
 in Relation to \\ Security \& Privacy Obligations 
}


\author{Christian Esposito}
\affiliation{%
  \institution{University of Salerno}
  \country{Italy}}
\email{esposito@unisa.it}

\author{Olaf Hartig}
\affiliation{%
  \institution{Link{\"o}ping University}
  \country{Sweden}}
\email{olaf.hartig@liu.se}

\author{Ross Horne}
\affiliation{%
 \institution{University of Luxembourg}
 \country{Luxembourg}}
\email{ross.horne@uni.lu}

\author{Chang Sun}
\affiliation{%
 \institution{University of Maastricht}
 \country{The Netherlands}}
\email{chang.sun@maastrichtuniversity.nl}

\begin{abstract}
The Solid specification aims to empower data subjects by giving them direct access control over their data across multiple applications. As governments are manifesting their interest in this framework for citizen empowerment and e-government services, security and privacy represent pivotal issues to be addressed.
By analyzing the relevant legislation, notably GDPR, and international standards, namely ISO/IEC 27001:2011 and 15408, we formulate the primary security and privacy requirements for such a framework.
  Furthermore, we survey the current Solid protocol specifications regarding how they cover the highlighted requirements, and draw attention to potential gaps between the specifications and requirements. We also point out the contribution of recent academic work presenting novel approaches to increase the security and privacy degree provided by the Solid project. This paper has a twofold contribution to improve user awareness of how Solid can help protect their data and to present possible future research lines on Solid security and privacy enhancements.

\end{abstract}




\maketitle

\section{Introduction}

The SOcial LInk Data (Solid) protocol~\cite{Sambra2016,Solid} is a draft specification for managing personal data 
to stimulate a drive for reusable W3C standards for privacy on the Web. 
An emerging key ambition of this project 
is to use Solid, in conjunction with legislation,
to empower Data Subjects regarding access to their own data, referred to as \textit{data sovereignty}.
Solid initially aimed to decentralise social networking, taking data out of the hands of corporations.
We focus however in this paper on the game-changing developments where companies and institutes 
propose to provide services based on Solid to store personal data intended for a variety of processing activities hence subject to EU General Data Protection Regulation (GDPR). 


Solid is currently evolving in a bottom-up fashion to support data sovereignty in GDPR-sensitive settings,
where instead of Processors storing the data from Data Subjects locally where the processing is performed, the data is stored by the Data Subjects and \textit{access} is given to the data via the Solid protocol. 
While a Processor may cache data for processing, subject to a \textit{legal basis}, it is obliged to log its accesses with the Data Subject through the Solid protocol, making data visible to the Data Subject.
Following GDPR, a Controller (typically the management of a company or organisation) defines and records the legal terms of the processing activities that Processors under their mandate may engage in.
Furthermore, the Controllers are obliged to respond to complaints from Data Subjects concerning Processors under their mandate, and may receive warnings and fines in the event of non-compliance from powerful Supervisory Authorities anointed by each EU member state.


While privacy is a stated goal of Solid, 
being compliant with current legislation on the topic is unattainable at the current status of the Solid specification. Even if a specification contains some protection means to build a system with privacy as a central requirement, some information might always be leaked to third parties, and, furthermore, the legal basis may be subject to conflicting and evolving legislation. As a practical example, Solid exploits OpenID Connect for identity management and authentication, which is a valid enabling technology, but it is also vulnerable to certain privacy issues~\cite{9229747,Hammann2020}.
OpenID Connect may also raise certain legal obligations in terms of data protection, since the identity provider can trace which relying parties the user logs in to -- which is a power given to the identity provider that is routinely monetised and quite opposed to the principle of data sovereignty~\cite{Jacobs2020}.
When making a privacy claim, several factors must be taken into account, including legal and ethical guideline drawn up by governments, e.g., EU General Data Protection Regulation (GDPR), and professional bodies, such as ISO27001 and Common Criteria.
We must take into account the relationship between various vulnerabilities in the system and also the good-practice management of a system~deployment.

The Solid ecosystem consists of distributed entities that provide infrastructure and identity services necessary to establish trust relations between various types of users.
Following their role according GDPR, the users can be further divided into roles, such as the Data Subject who typically owns the pod, and Processors who carry out processing activities for a particular purpose as defined by a Controller.
In this paper, we will expand on the above distributed system architecture that forms a vision for the Solid ecosystem, which we extract from a rapidly evolving suite of draft specifications. 
Taking care of privacy issues is all the more important for Solid, since, traditionally, knowledge graph technology, such as dereferenceable Linked Data, and Resource Description Framework (RDF) databases, were designed with open data in mind with the aim is to maximise access to data under suitable licenses. Thus if we directly reuse existing  technology, 
as the reference implementations of Solid do,
some of the guidelines and standard configurations are fundamentally opposed to privacy requirements.

The paper is structured as follows. Section~\ref{sec:back} describes the current status of the Solid ecosystem, its provided security and privacy solutions and a model of user responsibilities.
Section~\ref{sec:requirements} highlights the set of requirements coming from the main and widely recognised standards and legislation within the context of security and privacy. Section~\ref{sec:assessment} provides a coverage analysis of these requirements by reference to the current status of Solid project, thereby highlighting potential gaps between current specifications and privacy requirements.
The last section summarises 
potential future work that may achieve better coverage of the highlighted requirements.

\section{Background on Solid}
\label{sec:back}



Solid~\cite{Sambra2016} is a specification defining containers of personal data, called \textit{data pods}, and a set of open interfaces leveraging existing W3C standards and protocols.
Multiple \textit{pod providers} such as Inrupt.net or Solidcommunity.net provide infrastructure for hosting and managing pods, allowing the owner of a data pod to choose their own pod provider and also to migrate between providers, avoiding data lock-in~\cite{Abovetheclouds}. Access to data within a data pod by \textit{Solid apps} is standardised by the \textit{Solid protocol}, which defines authentication mechanisms, access control policies, and a RESTful API.


Providers act similarly to cloud storage services by
determining the internal rules for handling 
resources, and for maintenance, such as patching on-the-fly.
The data storage is organised in containers hosting RDF data and other documents. 
Data in pods may be written by the client of the user who owns the pod, as well as applications that are compatible with Solid (Solid apps).
A Solid pod is equipped with libraries to define and enforce access control policies and builds on the Linked Data Platform (LDP)~\cite{LDP}.
Pods offer communication means via the RESTful API, and also by leveraging notifications when their internal state is changed so that listeners can be informed of these changes.
Pods can be directly contacted by users through their client to request and obtain content and/or to insert/update data from any compatible provider. Pods may also be contacted indirectly, such as through an intermediary, that draws data from a network of pods in order to fulfil a service satisfying a user request.


Solid is no longer an academic project~\cite{mansour2016demonstration}, since pilots in the UK and in Flanders leverage it in e-governement service to 
put citizens in control of their own data~\cite{verbrugge_fitce_2021}.
Such applications place security and privacy demands on the Solid~project.
In Solid, identities are expressed  using WebID~\cite{faisca2016decentralized}, denoted by a URI that names an agent on the Web.
  Following Linked Data principles, if a WebID is looked up, a profile document (namely a WebID-Profile) is obtained where its referent is described.
  WebID-based protocols~\cite{oraskari2017access}, such as Solid OpenID Connect (Solid OIDC), WebID-TLS, WebID-TLS+Delegation,
  leverage this mechanism to realise  different approaches to log in.
Solid-OIDC is currently prevalent, where 
a user/service refers to another entity in the network (an Identity Manager, IdM) which can vouch for it. 
Authentication and authorisation protocols are defined by the OpenID Connect standard~\cite{sakimura2014openid}, which defines how a client and pod interact with the trusted third-party IdM.
This alleviates some security challenges for various sites since they need not handle their own ad-hoc login logic nor store authentication attributes, 
and users do not hold separate identities and passwords across multiple sites.
 The Web Access Control (WAC) specification~\cite{sacco2011access} provides an ontology for describing data accesses authorisations, and is to be superseded by an Access Control Policy (ACP) specification~\cite{ACP}. 
 A view of WAC, or ACP, as Access Control Lists (ACLs) for presenting the current access rights of a client with respect to a resource is presented in HTTP Allow headers.

Clients and servers in the Solid ecosystem communicate using HTTPS,
secured by digital certificates managed by an IdM, 
where the IdM may be federated~\cite{sanchez2012enhancing}.
On privacy, the specification is more vague, stating that providers are not encouraged to maintain identifiable information, and suggesting that users have control of their data by managing directly the ACL using WAC.
The intention is that Data Subjects, who store their data in their own pod, can determine for themselves whether an entity requesting to access the data has a \textit{legal basis} to use the data in a particular way.
For example, \textit{consent} and \textit{legitimate interest} are a legal bases, where the latter is usually a decision made by the company internally when the company that stores personal data directly.

Simply adopting these standards is not sufficient to cover
all security and privacy requirements and obligations of government-level data processing.
Currently, using Solid to manage personal data does not in itself mean that we have compliance with GDPR; rather, Solid shifts obligations related to the personal data privacy to the various entities within the Solid ecosystem, such as the \textit{pod user}, the \textit{pod provider}, and the providers and users of \textit{Solid apps}, as well as organisations associated with data, apps and users.
We argue that privacy concerns should not be left to each Pod provider, but should be enshrined in the specification of the Solid protocol that Pod providers implement.



Since Solid is not a centralised system where an authority may globally analyse and  manage flows of information throughout the system, we run into challenging problems such as managing trust and managing information leakage via indirect unintentional or maliciously manufactured flows of partial information.
Users able to run a client for a Solid app through which they have access to the pod, we divide into different actors. 
 The \textit{Data Subject} is the person about whom data is stored in a Solid \textit{pod}. 
 It can specify the access rules to be enforced within the pod, giving them permission to directly approve requests from others.
  The \textit{Data Consumer} is a user 
  that accesses the pod to request resources.
 They must be properly authenticated, and be authorised based on the policies set by, or in coordination with, the Data Subject.
The Data Consumers may be further classified as \textit{viewers}, which only read the data in a manner that is not subject to GDPR, e.g., sharing holiday photos with friends or internal exchanges within a company, and \textit{Processors}, which access data for automatic processing and potentially for decision making, hence are subject to GDPR. The distinction between Processors and viewers is motivated by Article~22 of GDPR -- a Data Subject may agree to let viewers read his/her data while forbidding Processors to perform automated decision-making. Data Processors are delegated the power to access data by a \textit{Controller} according to a specific legal base, which is not only the direct, free and explicit consent of the Data Subject.
Processors, must properly reflect the purpose of processing, such as a doctor treating a patient in an emergency or a researcher collating analytic. According to Article~35, the Controller has to perform a qualitative and preliminary risk assessment, to the rights and freedoms of Data Subject, and put in place all the measures to reduce the identified risks. For large and state organisations in the EU, a \textit{Data Protection Officer} (DPO) will be appointed to handle the attributes of Processors related to the organisation and to manage the privacy policies for the Controller. 
In addition to Data Consumers, there are also Data Feeders which are entities entitled to insert new data within the pod. Article~13 of GDPR indicates the legal obligations to be guaranteed when collecting personal data related to the Data Subject, and having this stored in a pod managed by the subject itself simplified the compliance to these obligations. 
Protocols between the Controller, Processor and Data Subject facilitating compliance with GDPR are not yet part of any Solid specification. 
Moreover, GDPR encompasses legitimate interest as one of the legal bases for the processing of personal data, but Solid does not include this possibility as the subject has always to explicitly allow processing actions for his/her pods.

A party that does not necessarily have access to the data inside the pod, is a \textit{Pod Provider} (excluding self-hosting options). They provide the infrastructure in which the data is stored and employ a \textit{Chief Information Security Officer} (CISO) to oversee the configuration and maintenance of its machines 
 running the servers
 containing the pods of the provider.
 The same view applies even if the provider outsources its infrastructure to the Cloud. 
    Cyber security standards mandate a CISO should have access to data logs in order to respond to technical incidents.
A Pod Provider is obliged by GDPR to assign a DPO whose responsibilities are expressed in Article 39, consisting of monitoring, documenting and reporting the compliance of the ICT solution against the legal obligations of GDPR and replying to possible inquires of the Data Subject, such as inquiries concerning the proper handling of their personal data.
 The DPO has access to the logs of the server and the ICT infrastructure, but,
 as the DPO is not meant to have a solid CS background, he/she may interact with the CISO, and does not have the same front-end.

There may be \textit{Solid apps} that act as \textit{intermediaries} between a Processor and Data Subject{~\cite{eisenstadt2020covid,sun2022}}.  For example, a Processor may indirectly access the data of Data Subjects via an intermediary Solid app that reads data on behalf of the client, and anonymises it before it reaches the Processor.
Indeed such intermediaries are Processors, with their own Controller, that have been delegated to access data on behalf of another Processor, and are required by GDPR to 
reflect the purpose of the access and credentials of the Processor(s) concerned, and to record the degree of anonymisation applied for auditing purposes.
Such delegations should be reflected when authenticating the intermediary, which may beyond the capabilities of existing IdMs.
Besides authentication duties, IdMs may also be involved in auditing, for example in the case of disputes where the identity of the Processors recorded in a cryptographic proof of accesses is verified.

\begin{table*}[h!]
\footnotesize
\begin{tabular}{|m{6mm}|m{71mm}|m{71mm}|c|c|}
\hline
\multicolumn{1}{|c|}{\textbf{Id}} & \multicolumn{1}{c|}{\textbf{Description}} & \multicolumn{1}{c|}{\textbf{Security Controls}}& \multicolumn{1}{c|}{\textbf{ISO}}& \multicolumn{1}{c|}{\textbf{GDPR}} \\ \hline
Req\_01 & Ensure access only to authenticated and authorized users and prevent unauthorized accesses  & A proper access control must have least privilege, revocation policies, and separation of duties across different roles & Yes & Yes  \\ \hline
Req\_02 & Guarantee correct and effective use of cryptography to protect the confidentiality, authenticity and/or integrity of data when stored and/or exchanged  & Sensible data exchanged between clients and server have to be properly encrypted, as well as when stored within the hosting machines.  & Yes & No  \\ \hline
Req\_03  & Protect the physical access to computing and communication equipment  & Secure deployment of the solution is needed to cope with internal attacks. & Yes & No \\ \hline
Req\_04  & Protect the configuration parameters, including identities, security claims and authorization policies, and proper management of users  & Secure with cryptography the configuration parameters, and defining a user management with no vulnerabilities and with a proper revocation mechanism. & Yes & Yes \\ \hline
Req\_05 & Data stored and managed by the system need to be analysed to determine if classifiable as personal and to characterize the legal base for their treatment & Risk assessment in the form of DPIA is needed, and one of the 6 legal bases within GDPR for processing personal data must be identified and enforced & No & Yes \\ \hline
Req\_06 & Collection and protection of event logs, for post-mortem analysis of the system behaviour and user activities  & Logging user activities, exceptions, malfunctions and events relating to information security should be carried out, maintained and periodically reviewed. & Yes & Yes \\ \hline
Req\_07 & Users are allowed to read their collected logging data, while only the DPO is entitled to read completely all the logs, which needs to be tamper-proof  & A pseudonym scheme has been implemented so that users are not traceable throughout the logs, and logs are stored on a tamper-proof DB. & No & Yes \\ \hline
Req\_08 & A proper login procedure needs to be put in place at the first use of the solution, by having a strong authentication scheme  & The used authentication should be more robust than a naive password-based one & Yes & Yes \\ \hline
Req\_09 & Unlinkability and unobservability  of the accesses to personal identities & Pseudo-anonimity of the used authentication and authorization tokens & No & Yes \\ \hline
Req\_10 & The rights expressed in articles 16, 17, 18, 21 and 22 must be enforced  & Advanced data management functionalities should be introduced and enforced by implementing a proper Privacy Information Management System in addition to a mere event logs & Yes & Yes \\ \hline
Req\_11   & Notification of data breaches  & Proper means should be provided. & No & Yes \\ \hline
\end{tabular}
\caption{Security and privacy requirements obtained from ISO/IEC 27001:2017 and the GDPR. 
}
\vspace{-5mm}
\label{tab:req}
\end{table*}

\section{Security \& Privacy Requirements}
\label{sec:requirements}

Security is related to the protection from potential harm intentionally caused by others by exploiting vulnerabilities. 
Privacy, on the contrary, is related to preserving data in its intended context, free from interference or intrusion from outside that context~\cite{Nissenbaum2004privacy}.
A secured system does not necessary cover privacy demands~\cite{schaar2010privacy}. 
Therefore, it is crucial to have a clear view of important security \& privacy requirements a system must support.

We summarise in this section high-level security \& privacy requirements expected of an information system handling personal data, relevant to Solid, from relevant standard and legislation, such as the ISO/IEC 27001:2017~\cite{din2017iso} and GDPR~\cite{calder2016eu}.
GDPR is complemented by the ePrivacy Directive~\cite{bond2012eu} which states in  Article 3 how it applies to the the provision of publicly available electronic communications services in public communications networks,
and also the upcoming ePrivacy regulation. 
Finally, also the ISO/IEC 15408 standard~\cite{bao2013supporting} has been considered within our analysis as it specifies functional and assurance requirements for computer systems, and also
ISO/IEC 27701\cite{lachaud2020iso} that presents, in Annex D, a mapping between the introduced ISO controls and the ones in the GDPR.
In applying GDPR, organisations are faced with the free interpretation of operating instructions and multiple pragmatic aspects, while the ISO 27001 standard integrates with Article 32 of the GDPR as it defines best practices for mitigating the risks within the organisation by extending the Information Security Management System from ISO/IEC 27001 to privacy management.

ICT services are vulnerable to various kinds of external attacks that adversaries can trigger by exploiting the vulnerabilities within the network or the publicly-accessible API. The intent of these attacks is to compromise one or more security or privacy properties.
Proactive countermeasures include detecting and remove vulnerabilities that can be exploited to conduct attacks from both the specifications and implementations,
while reactive countermeasures include detecting possible misbehaviour at runtime, limiting damage. These can be extracted from the controls within the ISO/IEC 27001:2017 or the Common Criteria within ISO/IEC 15408.
 To avoid possible Denial-of-Service attacks, access control and anomaly detection should be applied.
Man-in-the-Middle attacks can be reduced by continually reviewing notifications about vulnerabilities in relevant libraries and standards.

Services holding personal data (such as pods and processors) or monitoring user activities (such as the IdMs, intermediaries, or the CISO) 
may be subject to professional standards such as Common Criteria (CC)~\cite{CC2017}.
The privacy property \textit{unobservability} in CC can be realised if a user interaction with the system happens without others, especially third parties, being aware that a specific data instance or operation has been used by a given identifiable entity.
Solid makes unobservability challenging, since some communications which might in traditional information systems have been internal to an organisation are exposed over the Web, and hence AI-based attacks can infer surprisingly precise information from HTTPS-based APIs while knowing only connection timings and packet sizes~\cite{Chen2010}.
A weaker property defined by CC is \textit{pseudo-anonymity}, where users interact with the system without disclosing their identity, but can still be accountable later if obliged by law to trace past activities in the case of a dispute. 
A relevant property  is \textit{unlinkability}, where  users may make multiple uses of resources or
services over a time frame, without others being able to link these uses together.

In principle, proper mechanisms need to be implemented so that the requirements in Table~\ref{tab:req}, inferred from the ISO/IEC 27001:2017, GDPR and ePR, are in a well-informed balance.
We have indicated clearly whether a requirement comes from security guidelines or from legal privacy obligations. Compromises may be required when requirements are in conflict, e.g., while security demands logs, the Data Subject has the right to be forgotten.





\section{Security \& Privacy Assessment}
\label{sec:assessment}

This section presents systematical assessments on the degree of satisfaction for security and privacy requirements (Table~\ref{tab:req}) with respect to the current Solid specification and the implementations. We explain to what extent it is covered by the specification and concrete implementations; and where the potential vulnerabilities or weaknesses in the system. We identify gaps in the current specifications that may be addressed by evolving the Solid specifications.

\subsection{Ensure access to authorised users (Req\_01)}

Authentication for Solid is facilitated mainly by Solid OpenID Connect (Solid-OIDC)~\cite{OIDC,OpenIDC}.
OpenID Connect is widely-deployed with robust libraries, but there remain vulnerabilities that may be addressed in the specification. 
Beyond the scope of OpenID Connect, Solid-OIDC allows clients with no previous trust relationship or identity to assume a generic identity.

A weakness of OpenID Connect is that the IdM of the Data Subject can learn behaviours of the Data Subject connecting to their own Pod.
This may be avoided by using an anonymous credential system, such as IRMA~\cite{Jacobs2020}, where the identity provider signs the attributes for authentication that prove the Data Subject owns the pod.
The W3C is taking steps in this direction via DIDs and Verifiable Credentials~\cite{DID2022,Verifiable2022}. However, a verifiable credential-based authentication mechanism has not yet been established for Solid.

Exposing information flows concerning the CISO to an IdM may reveal information about the business model to the pod provider. 
However, the main point of concern should be revealing the behaviour of Processors.
Since an IdM can learn over the time which Processors access a Pod,
this may expose sensitive information about the processing activities a Data Subject is subject to, thereby impacting the privacy of the Data Subject.
The anonymous authentication mode, as the sole extension to Solid-OIDC allowing clients to connect anonymously, does not resolve this. Not least because we should authenticate at least a credential confirming that a Processor has been approved to access data, even if uniquely identifying information is not revealed. Hence, fully anonymous access is a problem for the accountability principle of GDPR, as the pod cannot verify the credentials of who is accessing certain data, nor can identities be revealed retrospectively by the IdM, if required to resolve disputes. 
Verifiable credentials map well to GDPR, since the Controller would act as an \textit{Issuer} of credentials to Processors they are responsible for. Upon access, the Processor presents their issued credential anonymously to the data pod when authenticating. 
This protects the identity of the Processor while also preserving accountability.

Another problem related to OIDC is the lack of built-in support for explicit Consent by Data Subject to express who is allowed to access certain data, for which purpose and for how long (see also Req\_05). The available semantics of the access control in OIDC is not able to express such a complex set of policies. This is a known by the Solid community, and is behind a shift in the specification from ACL to Access Control Policies (ACP) with richer policies
which can be combined with legal vocabularies (see Req\_05).

There are some vulnerabilities and limitations in OIDC that developer should be aware of~\cite{Mitchell2016, Fett2016}.
We review three known vulnerabilities below, and describe how they impact Solid, and how they may be addressed by tightening the specification of Solid-OIDC.

\textit{Malicious Pods.} A malicious data pod may gather data about clients to hijack their identity and access rights.
This is a viable threat model for Solid, since anyone is permitted to host their own infrastructure for their data pod.
Having received credentials from a user, the \textit{identity provider} (i.e., OAuth server), must never use a HTTP 307 (Temporary Redirect) status code. 307 redirect replays the credentials in the body of the POST to the data pods, which are not guaranteed to handle a client's credentials. Instead, the identity provider MUST use a HTTP 303 (See Other) status code while receiving credentials from a user. 

\textit{IdM hijacking.} Even if the Pod, client and Idm are honest,
an an attacker may authenticate themselves to access the data pod with the access rights of the honest client, as follows. The attacker tricks the pod into using a malicious IdM controlled by the attacker to authenticate the honest client who does not use that IdM.
The attacker manipulates the first message so that the client directs to their honest IdM for the actual authentication. Since the pod thinks the attacker's IdM was used, while issuing access tokens, the pod unwittingly issues the access token to the attacker.
This attack can be mitigated, by the identity provider recording its own identity, when the client is redirected back to the data pod.

\textit{Session hijacking.}  
Assuming again the Pod, IdM, and client are honest, but a URI inside the resource requested may refer to something controlled by the attacker.
Upon a redirect, due to the honest client selecting the URI, the attacker learns the state information from the URI from which it was redirected, which appears in the header of the redirect. This is a seed for various attacks to learn the state information. This can be mitigated by the pod implementing a \textit{referrer policy} that instructs the browser to strip away the state information from the referrer field of the header, which is perhaps best considered as a clarification of the Cross-Origin Resource Sharing recommendation in the draft Solid specification~\cite{Solid}.

\subsection{Effective use of cryptography (Req\_02)}

The use of HTTPS, 
in a RESTful API for sharing private data between data pods and authenticated/authorised clients,
does not necessarily guarantee privacy; hence we examine how effectively HTTPS is employed in the Solid. 
We focus here on which aspects of the RESTful API of Solid should be addressed, due to privacy issues.

Consider an anonymity scenario where a Data Subject makes health record available to a doctor but not to 3rd parties (even the existence of the data).
Assume personal data of a citizen lies behind an authentication mechanism such as Solid-OIDC,
and that an adequate trust framework exists so that the citizen may verify the credentials of health specialists.
Also assume the data is made available via the HTTPS URI: 
\textit{https://john.provider.net/vaccinationdata.ttl}.
There is a crude but effective attack on privacy where the attacker poses as an App trying to access a resource and observes from the HTTP response whether a resource exists. Suppose that 404 NOT FOUND is the response for resources that do not exist
and 403 FORBIDDEN is for when a resource exists but the App is not authenticated to access to. Then, an attacker can determine that a Data Subject has a vaccination record even though the attacker has no access to the record, thereby violating privacy.
The Solid specification addresses the leakage of information via REST by suggesting 
return a 404 NOT FOUND response to all Solid clients that are not authenticated and authorised for the resource to prevent trivial probing attacks.
However, this is undermined by one requirement in the Solid draft at the time of writing that ``servers MUST respond with the 405 status code to requests using HTTP methods that are not supported by the target resource.'' This and other conflicts in requirements should be resolved by prioritising the 404 NOT FOUND status code for unauthenticated or unauthorised clients.

Another issue is that the specification permits HTTP URIs. HTTP URIs are redirected to their https counterparts using a 301 status code and a location header. The specification states that ``a data pod SHOULD use TLS connections through the HTTPS URI scheme to secure the communication between clients and servers.'' However, ``SHOULD'' should be upgraded to ``MUST'', since the protocol for handling HTTP URIs reveals the URI that is being requested to any eavesdropper.
In addition, the specification should be explicit about the version of TLS used. The most widely deployed protocol TLS 1.2 has known vulnerabilities~\cite{Raccoon}. While the use of TLS 1.3 may cause incompatibility with Apps that do not implement the latest standard~\cite{TLS3}. 
TLS 1.3 also has privacy flaws, for example, stemming from revealing the identity of the server in plaintext~\cite{Karthik}. However, addressing such issues is beyond the scope of Solid. 

Regarding effective cryptography other than HTTPS, GDPR stipulates cryptography to improve trust in the access logs, which is not currently covered by the Solid protocol (see also Req\_06). 
Moreover, Solid does not mandate the use of cryptography when storing data in pods (see also Req\_03).

\subsection{Protect Physical Access (Req\_03)}

Physical protection is the concern of the management policy of \textit{pod providers}. 
However, to promote trust in pod providers, it would be advisable for specification in the Solid ecosystem to mandate mechanisms protecting against insider threats.
The storage of servers should be encrypted reducing risks associated with attackers with physical access to servers.
There should be a transparent policy regarding who has access to the encryption keys to disks, and under what conditions.
Trust in pod providers can also be established by being transparent about the status of the provider with respect to security audits, which could be important aspect of governance of the Solid ecosystem.
Provisions should be made for transferring pods between providers securely. This requires physical and network security, and standardisation of protocols for handover. The current specification has not covered inter-provider exchange of pods.

\subsection{Protection of Identities and Policies (Req\_04)}

Client details and access control policies are stored in a database and can be queried when an authentication and authorisation decision is needed.
GDPR specifies that when the digital identity of the Data Subject is used in data processing, it should be tied to their authentication credentials.
There is a mismatch between GDPR and Solid here, since the Processor is authenticated to access the data stored by the Data Subject.
This leaves the question of how the Processor authenticates the Data Subject, particularly if there are no prior trust relationship. For instance, a doctor accessing a pod of a new patient may be tricked into using the pod of another patient, without some means of authenticating the Data Subject associated with a pod. One potential solution is to use attributes from the Data Subjects such as registered locations or verified affiliations. However, this issue has not been addressed in the current Solid specification.

\subsection{Identification of Eventual Legal Basis (Req\_05)}

The Access Control Policy 
should take into account for what purpose the personal data of a Data Subject is being processed.
That is, a policy should explicitly reflect the legal basis for a Processor accessing data in a pod, as recorded by their respective Controller.
There has been work on explicitly recording the purpose of a data as part of the access control logs~\cite{Havur2020,Esteves2021,Esteves2022,Debackere2022}.
An authorisation request should record the reason why the access is requested, explaining what the user will do with the data. For example, the data may be used once in a computation to find the percentage of recovered COVID patients in a community, without revealing an individual's status. 
For \textit{data retention} purposes, which is another requirement of GDPR, it would be beneficial if the policy stipulates whether the data must be destroyed after use by the Solid app, or whether the data is permitted to be used only within a particular time window.
The Data Protection Vocabulary is a key step towards presenting such structured information, so that the legal basis may be taken into account by automated agents.
Such steps are ``encouraged'' by GDPR (cf.\ Recital 100),
``allowing data subjects to quickly assess the level of data protection of relevant products and services.''

A question here is who is responsible for identifying the eventual legal basis for personal data.
The vast majority of Data Subjects will not have the legal expertise to make a judgement about what legal basis applies in a scenario.
Ultimately, the responsibility for identifying a legal basis is the Controller responsible for the Processor
providing and processing the data placed in a pod.
Thus responsibility for a failure to record the correct legal basis for access by a Processor 
at the time of each authorisation request lies with the Controller,
whether or not the Processor followed correctly the guidelines recorded by their Controller.
Thus, by a pod requiring that the Processor indicates their legal basis at the time of an authorisation request,
the Data Subject has stronger grounds for holding the Processor accountable, via their Controller, in the case of a dispute.

\subsection{Event Logs (Req\_06 and Req\_07)}

Logging is a key mechanisms according to ISO 27001 for obtaining snapshots of a system in order to detect possible sources of breaches and to improve security policies and solutions in place.
In addition, in accordance with GDPR, logs may provide evidence in the event that concerns about access are raised (see also Req\_10). 
Indeed, we argue accesses to data are more pertinent than the data itself with respect to GDPR, since the access logs reveal information about how personal data is being used.
Thus logs should be tamper proof either by cryptographic means or by use of a trusted third party,
so as to be valid for use in a court of law.
Also a proper authentication and authorisation solution should be put in place to discriminate who can have access to what kind of logs. 
A candidate trusted third party could be the pod provider, perhaps operating a database for logging that the Data Subject cannot manipulate.
Currently, the Solid specification does not define an effective logging system, but we believe it is a priority. Records in logs can build on ACP~\cite{ACP}, which can describe access instances using \textit{context graphs}, recording for example the Processor, the resources accessed, operations performed, the agreed terms of the policy granting access.

A view of access logs presented to CISO may apply pseudo-anonymisation, perhaps via an intermediary for multiple pods; and, although a pseudonym scheme is not sufficient to fully anonymise logs since there are numerous deanonymisation attacks, a CISO should be trusted not to perform such attacks.
At minimum, in order to detect technical incidents, such DDoS attacks, a CISO may have access only to the logs of the underlying system, concerning IP addresses of connecting clients for instance, but does not have access to the access logs internal to Solid.

\subsection{Data Protection Officer and Logs (Req\_07)}

Article~22 of GDPR requires the designation of the Processor's activities to be recorded by the Controller.
For large organisations and public entities the Controller assigns a DPO for whom recording the activities of Processors is one of their formal duties.
For example, the DPO of a university records in an internal information system the legal basis for processing activities related to each ongoing research project.
For each processing activity, the Controller should record their own contact details and also the following information:
the purposes of the processing (c.f.~the data protection ontology {called Data Privacy Vocabulary}(DPV)~\cite{DPV}),
the categories of Data Subjects and of the categories of personal data (e.g.~specific pods, URIs, shapes),
and the time limits for erasure.
The Controller is also obliged to have a record the categories of recipients to whom personal data will be disclosed.
The Controller also details security measures including:
pseudonymisation and encryption of personal data,
the ability to ensure the ongoing confidentiality, integrity, availability and resilience of processing systems and services;
the ability to restore the availability in the event a technical incident;
provision organisational measures continual security evaluation.
Such concepts are becoming provisioned for through the DPV, making formal representations of these obligations internal to the Solid ecosystem a real prospect.

The Processor must also maintain records of all their processing activities on behalf of a Controller.
This includes details on any delegations on behalf of another Processor, along with their respective Controllers.
Delegations may be applicable, for example, when the processor provides a service such as caching or a privacy-preserving service that is offered to another Processor.
In addition to the above delegations of authority,
the  Processor records: categories of Processing,
and a general description of the technical and organisational security measures.
Notably this information is also recorded by the Controller. 
This supports our view that the Controller is ideally placed to act as an Issuing Authority in an anonymous credential system, which certifies the above credentials of the Processor, which the Processor can then use directly. 
While the Data Subject may record all relevant information about accesses in the logs of their pod,
the Controller and Processor are in fact the only agents obliged by GDPR to maintain logs.
The view of the logs maintained in the pod is more so and enhancement of transparency rather than an obligation under GDPR.
It avoids the need for a Data Subject to formally request the information, which can lead to lenthy delays and unsatisfactory incomplete information pertaining to accesses provided by the Processor by post for example. 

There are no interfaces in Solid between the DPO of the Controller, and other entities.
Related work proposes that the Controller engages in a protocol with the Data Subject~\cite{Debackere2022} when involving  \textit{consent}.
A limitation of engaging the Controller when authorising Consent is that a DPO typically does not have resources to respond manually to all the requests.
There is a case to extend protocols to cover the legal bases for all data processing scenarios, such as Legitimate Interest, and to standardise such protocols to facilitate communication between the Controller, regardless of the information system employed by the DPO of the Controller.

\subsection{Multi-Factor Authentication (Req\_08)}

User account management is well-known to be one of the primary sources of data breaches.
Multi-factor authentication is particularly important for account initialisation, account recovery, and for critical actions such as accessing the access control logs by a CISO. Multi-factor authentication is not recommended in Solid-OIDC yet, however
including it explicitly
ould increase confidence in Solid. 
If a Data Subject grants access to a Processor with whom the Data Subject has never had interaction with, then multiple factors such as proximity may be authenticated.

\subsection{Unlinkability to Personal Identities (Req\_09)}

In scope of Solid is the privacy property \textit{unobservability}, which ensures that certain operations cannot be observed by a man-in-the-middle.
If some functionalities are outside the pod, 
such as the Wallet containing logs proposed in related work~\cite{Debackere2022},
there is the unnecessary risk of exposing security-critical data over the internet. 
Thus critial access is better made internal to the pod, hence unobservable with respect to man-in-the-middle attacks on the network.
In a data sovereignty scenario involving a pod controlled by the Data Subject, rather than the Processor,
there is no possibility of 
making all operations unobservable.
When we do not have unobservability, the strongest privacy property we can target is unlinkability.
For Solid-OIDC, this means that a man-in-the-middle cannot connect two authentication
attempts by the same User, be it a Data Subject,
a Processor, or some other Agent. 
Standard OpenID Connect provides no provisions for unlinkability.
This issue has been identified, and potential solutions such as 
Privacy-Preserving modification to OpenID Connect~\cite{Hammann2020}
or new single-sign-on protocol such as in~\cite{Fett2015}.

Timing side-channels can 
violate unlinkability, since the slightest variation in timing between different information flows can reveal information that can be used by malicious third parties to profile Processors and Data Subjects. 
The Solid draft specification mentions protecting against timing attacks, without being specific.
HTTPS traffic between pods and apps can reveal significant information,
inferring information about the resources accessed.
An attacker can trivially infer the fact that \url{john.provider-name.net} is contacted by a particular App,
hence the domain should not reveal the name of the data subject as a sub-domain, as is the case for Inrupt for example.
To reinforce this consider also that a man-in-the-middle knows the following
the length of the URL being accessed, rounded up to the nearest block length in the cipher, say 128bits;
The length of the resource in the response;
and the response time between initiation of the session and termination of the session.
Indeed, in reasonably busy applications, such as those with a dynamic Ajax API such information has been shown to be reveal fine-grained information, such as keywords being typed~\cite{Chen2010}. 
Even coarse-grained information inferred in this way can be used to trace behaviours profiling a Data Subject. 
Avoiding such profiling attacks would require a radical redesign of the Solid protocol, perhaps
mandating intermediaries
that obfuscate the above information.

\subsection{Respect the Rights of Data Subjects (Req\_10)}

Articles 16--18 of GDPR
enshrine the right of Data Subjects
to rectify their personal data,
erase their data entirely or 
restrict its usage for data processing.
By Article~19, the 
Controller has the responsibility
to notify the Data Subject of compliance
with such~requests.
In Solid, Data Subjects always have control of the data in their own pod which means they are always able to write, modify, and remove data elements and files in their pods.
There are however scenarios where a contractual agreement
does not allow the Data Subject to modify, remove
or restrict data stored in their own pod freely,
even if they have full power to inspect the 
data, access control policies and logs.
One such scenario may be if data is required for billing purposes or medical records.
Alternatively, a Data Subject may lack the time or expertise.
In addition, the right to restrict processing
should be triggered via Article~19 and not
unilaterally by the Data Subject.
A protocol between the Data Subject, Controller, and Processor would facilitate compliance with Article~19, perhaps with a pod provider acting as a broker.
Article 20 enshrines the right to data portability,
allowing the Data Subject to obtain their personal data in a structured format.
Thus Article~20 is undoubtedly catered for fully by Solid.
Article~21 ensures a Data Subject may challenge 
whether there is a legitimate basis for processing.
This is partly catered for by Solid, in that 
there is transparency about the data stored and 
the accesses.
However, a protocol could facilitate the notifications to a Controller
about specific objections by the Data Subject,
and by the use of cryptography to ensure tamper-proof logs.

\subsection{Notification of Data Breaches (Req\_11)}

Both data access logs, recorded for GDPR purposes,
and underlying network and system logs can be
used by the pod provider to detect and assess
potential technical incidents.
There are currently no provisions for notification of data breaches in the Solid specification.
This requirement is an example of how policy can impact the attacker model. The law requires that certain logs are maintained and assessed for signs of the presence of certain attackers. Thus, for instance Solid pods that inject fake data into a network in order to act as Sybils, which facilitate reidentification attacks after anonymising data, may be detected.
By keeping the number of Sybils in a network low, some intermediary services preserving privacy using techniques such as $k$-anonymity and $\ell$-diversity can be more effective~\cite{Sybil}.

\section{Conclusion with Recommendation}

This paper goes beyond the security and privacy review, eliciting requirements from ISO standards and GDPR and featuring self-review questions that appears in the draft Solid protocol at the time of writing. 
We reflect on challenges uncovered for strengthening privacy.
Foremost, \emph{the access logs should be an integral part of a Solid pod}, as needed for security and privacy audits,
such evidences the Data Subject can use to  challenge Processors or proofs of occurrend incidents to be analysed post-mortem. 
Providing such services externally to a pod creates unnecessary risks as more privacy-critical interactions are exposed over the internet than required. Current low-level APIs for granting access leave room for developers,
who need not be privacy experts,
to make mistakes.
In logs, the purpose of an access and credentials of the Processor should be
recorded cryptographically in the logs, ensuring that the Controllers responsible for accesses violating their legitimate basis 
may be held accountable.

We also drew attention to the need to \emph{upgrade authentication},
e.g., making credentials anonymous but yet accountable (e.g., we know a scientist from a particular institution accessed the data, but we do not know who, unless the Controller is forced to reveal it in the case of a dispute).
Authentication should provide means for delegation of access (e.g., when a Processor delegates part of its processing to a privacy-preserving service operated by another one). Despite having Solid bounded to OIDC, it is valuable to investigate and identify pros and cons of other different and competing specs, such as W3C WebAuthn~\cite{klieme2020fidonuous} or the blockchain-based Self-sovereign identity (SSI) solutions~\cite{liu2020design}, when applied to Solid. 
Since the Controller is obliged by GDPR to record the most detailed information on processing, and is the contact point for disputes, \emph{protocols involving the Controller need to be explicitly defined}. Closely related to this, Solid should \emph{provide all the possible legal bases} listed within the GDPR, and not only limited to the explicit consent. How modelling and integrating them within the data management of Solid is a hugh open issue.
\emph{Proper means to protect data}, which may be personal w.r.t. users but also used in the authentication/authorization schemes, should be embedded within a pod instance, since only leveraging on HTTPS resolve only a portion of the problem. 
Finally, we emphasise that the cyber security of any system is a moving target, since new vulnerabilities are discovered in key standards and libraries on a regular basis. Some of the suggestions in this work (e.g., the guidelines for tightening OAuth 2.0 in Req\_01) illustrate this kind of evolution. A possible path for addressing this is to \emph{separate the core Solid protocol from an evolving Security \& Privacy review} that is updated as vulnerabilities are disclosed. This can serve as a policy benchmark for pod providers and developers to adhere to, thereby improving trust in the Solid ecosystem.


\balance

\bibliographystyle{ACM-Reference-Format}
\bibliography{privacy}

\end{document}